\documentclass[twocolumn,showpacs,amsmath,amssymb,prl]{revtex4}

\usepackage{graphicx}
\usepackage{dcolumn}
\usepackage{bm}

\begin{document}
\title{
Dynamics of localized spins coupled to \\
the conduction electrons with charge/spin currents
}
%
\author{Masaru Onoda$^{1,3}$}
\email{m.onoda@aist.go.jp}
\author{Naoto Nagaosa$^{1,2,3}$}
\email{nagaosa@appi.t.u-tokyo.ac.jp}
\affiliation{
$^1$Correlated Electron Research Center (CERC),
National Institute of Advanced Industrial Science and Technology (AIST),
Tsukuba Central 4, Tsukuba 305-8562, Japan\\
$^2$Department of Applied Physics, University of Tokyo, Bunkyo-ku, Tokyo 113-8656, Japan\\
$^3$CREST, Japan Science and Technology Corporation (JST), Saitama, 332-0012, Japan\\
}

\begin{abstract}
The effects of the charge/spin currents of conduction electrons 
on the dynamics of the localized spins are studied in terms of 
the perturbation in the exchange coupling $J_{K}$ between them.
The equations of motion for the localized spins are derived 
exactly up to $O(J_{K}^2)$, and the  
equations for the two-spin system is solved numerically.
It is found that the dynamics depends sensitively upon the relative
magnitude of the charge and spin currents, i.e., it shows 
steady state, periodic motion, and even chaotic behavior.
Extension to the multi-spin system and its
implications including possible  ``spin current detector'' are 
also discussed.
\end{abstract}
\maketitle
The interplay between the current and spins is the subject of 
recent intensive studies \cite{GMR,SpinValve,
Sloncewski,Berger,Bazaliy,Versluijs,
Allwood,Yamanouchi,Yamaguchi,SZhang,
TataraKohno,SZhang2,BarnesMaekawa,CMR}.
The giant magneto-resistive effect (GMR) \cite{GMR},
the spin valve effect \cite{SpinValve}, 
the current-driven domain wall motion
\cite{Sloncewski,Berger,Bazaliy,
Versluijs,Allwood,Yamanouchi,Yamaguchi,
SZhang,TataraKohno,SZhang2,BarnesMaekawa},
and the colossal magneto-resistance (CMR) \cite{CMR}
are the representative phenomena due to the interaction between
the spin state and the charge dynamics.
Especially the charge current in the ferromagnetic system
is necessarily spin polarized, and the spin torque
associated with it drives the nontrivial phenomena
\cite{Sloncewski,Berger,Bazaliy}. 
On the other hand, the pure spin current without the charge current
is highlighted in the studies of spintronics
(such as spin Hall effect \cite{MNZ,Sinova}
and quantum spin pumping (such as spin battery \cite{Tserkovnyak}).
Spin current is present even in 
the insulating magnets. In the latter case,
the electric polarization is induced by the spin current
associated with the non-collinear spin configuration,
offering a new mechanism of the magneto-electric effect \cite{Katsura}.

However, the direct observation of the spin current
is difficult because it does not appear in the Maxwell equations.
One possible way is to measure the electric field or 
voltage drop accompanied
by the spin current due to the Aharanov-Casher effect
\cite{TakahashiMaekawa,PZhang}.
However, the magnitude of the voltage drop is estimated to be very small.
Therefore a more sensitive method to detect the spin current is desirable,
which we propose below.

In this paper, we study the dynamics of the local spins 
weakly coupled to the conduction electrons with charge/spin currents.
The equation of motion for the localized spins are derived
up to the second order in the Kondo coupling $J_K$ between 
the conduction electron spins and localized spins. 
This equation of motion is solved numerically and variety of 
dynamics is found including the steady state, periodic motion, 
and even the chaotic motion
in the phase diagram of charge current $\bm{j}^{c}$
and spin current $\bm{j}^{s^{\mu}}$.
It is not unrealistic to control both of them by several 
methods proposed \cite{MNZ,Tserkovnyak} in addition to the 
conventional spin-polarized current injection from the ferromagnets.
It is also noted that the polarization of the spin current component 
is controllable only in the case of weak $J_K$.
In the strong coupling case, the spin of the conduction
electron is forced to be parallel to the localized spin
at each site, and we cannot arbitrarily manipulate its direction.
\\
\noindent
{\it Model and equation of motion}-- 
We start with the double-exchange model of continuum version
treated in the closed time-path Green's function (CTPGF) formalism
\cite{CTPGF},
which is recently applied to the spin dynamics
in a Josephson junction.
(We shall take the units in which $\hbar = c = 1$.)
\begin{eqnarray}
&&\mathcal{Z}[\bm{S}(\bm{r},t)] 
= \mathrm{Tr}[T_{p}\exp(i\mathcal{S})],
\\
&&\mathcal{S} = \mathcal{S}_{\mathrm{WZWN}}+\mathcal{S}_{\mathrm{ext}}
+\int_{p}dt\int d\bm{r}
\Biggl[
i c^{\dagger}(\bm{r},t)\partial_{t}c(\bm{r},t)
\nonumber\\
&&\quad\quad
-H_{0}
+\frac{v_{\mathrm{uc}}J_{K}}{2}\bm{S}(\bm{r},t)\cdot\bm{\sigma}(\bm{r},t)
\Biggr]
,
\\
&&\int_{p}dt
 = \int_{-\infty}^{\infty}dt_{+}-\int_{-\infty}^{\infty}dt_{-}
\end{eqnarray}
where $\mathcal{Z}[\bm{S}(\bm{r},t)]$ is the partition function 
in the CTPGF formalism,
$T_{p}$ is the time-ordering with respect to the closed time-path,
$\mathcal{S}_{\mathrm{WZWN}}$ is the Wess-Zumino-Witten-Novikov (WZWN) 
term describing the Berry phase of the localized spins,
$H_{0}$ is the Hamiltonian for free electrons.
For simplicity, we consider $H_{0}$ of the quadratic dispersion and
isotropic effective mass $m^{*}$.
$\mathcal{S}_{\mathrm{ext}}$ represents the coupling between the system
with the external electromagnetic field and 
the driving force for the spin current of the conduction electrons.
The effective action for $\bm{S}(\bm{r},t)$ is given 
by $\mathcal{S}_{\mathrm{eff}}=-i\ln\mathcal{Z}[\bm{S}(\bm{r},t)]$.
Note that the constraint $|\bm{S}(\bm{r},t)| = S = \mathrm{const.}$
is implicitly imposed.

Here we consider the weak coupling regime, $|J_{K}|\ll \epsilon_{F}$
($\epsilon_{F}$ : the Fermi energy)
and retain the terms up to the  quadratic order in $J_{K}$.
Then, as in Ref.\cite{Balatsky}, 
$\mathcal{S}_{\mathrm{eff}}$ is estimated as follows.
\begin{eqnarray}
\mathcal{S}_{\mathrm{eff}}
&\cong&
\mathcal{S}_{\mathrm{WZWN}}
+g\mu_{B}\int_{-\infty}^{\infty} dt \int d\bm{r}\bm{B}(\bm{r}, t)
\cdot \bm{S}_{\Delta}(\bm{r},t)
\nonumber\\
&&
-\left(\frac{v_{\mathrm{uc}}J_{K}}{2}\right)^{2}
\int_{-\infty}^{\infty} dt dt' \int d\bm{r}d\bm{r}'
\nonumber\\
&&\quad\times
S^{\mu}_{\Delta}(\bm{r},t)
\Pi^{R}_{\mu\nu}(\bm{r},t;\bm{r}',t')
S^{\nu}_{c}(\bm{r}',t')
,
\end{eqnarray}
where 
$\bm{S}_{c}(\bm{r},t) = [\bm{S}(\bm{r},t_{+})+\bm{S}(\bm{r},t_{-})]/2$,
$\bm{S}_{\Delta}(\bm{r},t) = \bm{S}(\bm{r},t_{+})-\bm{S}(\bm{r},t_{-})$,
and
$\Pi^{R}_{\mu\nu}(\bm{r},t;\bm{r}',t') =
-i\theta(t-t')
\langle[\sigma_{\mu}(\bm{r},t),\sigma_{\nu}(\bm{r}',t')\rangle_{J_{K}=0}$
is the retarded spin Green's function of the conduction electrons
in nonequilibrium, which represents the response of 
the spin polarization of the conduction electrons
to the localized spins in the linear order, i.e.,
\begin{eqnarray}
\langle\sigma_{\mu}(\bm{r},t)\rangle
&\cong& -\frac{v_{\mathrm{uc}}J_{K}}{2}\int_{-\infty}^{\infty} dt' \int d\bm{r}'
\Pi^{R}_{\mu\nu}(\bm{r},t;\bm{r}',t')S^{\nu}_{c}(\bm{r}',t').
\nonumber\\
\end{eqnarray}
The variational principle
$
\delta\mathcal{S}_{\mathrm{eff}}/\delta\bm{S}_{\Delta}(\bm{r},t)
=0
$ 
at $\bm{S}_{\Delta}(\bm{r},t)=0$
gives the equation of motion for localized spins,
\begin{eqnarray}
\frac{\partial\bm{S}(\bm{r},t)}{\partial t} 
&=& 
\bm{S}(\bm{r},t)\times
\left[g\mu_{B}\bm{B}(\bm{r},t)
+\frac{v_{\mathrm{uc}}J_{K}}{2}\langle\bm{\sigma}(\bm{r},t)\rangle
\right],
\nonumber\\
\end{eqnarray}
where we have used the abbreviation 
$\bm{S}_{c}(\bm{r},t)\to \bm{S}(\bm{r},t)$.

For simplicity, we shall consider the case $\bm{B}(\bm{r},t) =0$ hereafter.
In order to estimate $\Pi^{R}_{\mu\nu}(\bm{r},t;\bm{r}',t')$,
we approximate the anti-commutation relation 
$\{c_{\sigma}(\bm{r},t),c^{\dagger}_{\sigma'}(\bm{r}',t')\}_{J_{K}=0}$
by its c-number value in equilibrium,
and the nonequilibrium nature of the system
is taken into account through 
the nonequilibrium distribution function $g(\epsilon_{k})$,
where $\epsilon_{k} = k^2/(2m)$.
\begin{eqnarray}
&&\langle 
c^{\dagger}_{\sigma'}(\bm{r}',t')c_{\sigma}(\bm{r},t)
\rangle_{J_{K}=0}
\nonumber\\
&&\cong
\int \frac{d\bm{k}}{(2\pi)^3}
e^{i\bm{k}\cdot(\bm{r}-\bm{r}')-i(\epsilon_{k}-\mu)(t-t')}
g_{\sigma\sigma'}(\epsilon_{k}).
\end{eqnarray}
Here we adopt the following form of $g(\epsilon_{k})$ 
in order to express the state with 
the charge current $\bm{j}^{c}$ and the spin current $\bm{j}^{s^{\mu}}$
of the conduction electrons.
\begin{eqnarray}
g(\epsilon_{k})
&=& f(\epsilon_{k}) 
+
\left[e\tau\bm{v}_{\bm{k}}\cdot\bm{E}
+e_{s}\tau\bm{v}_{\bm{k}}\cdot\bm{E}^{s^{\mu}}\sigma_{\mu}
\right]
\left[-\frac{df}{dE}(\epsilon_{k})\right],
\nonumber\\
\end{eqnarray}
where $f(\epsilon_{k})$ is the distribution function in equilibrium,
$\tau$ is the electron lifetime,
$\bm{v}_{\bm{k}}$ is the group velocity,
$\bm{E}$ is the external electric field,
$e_{s} = 1/2$ and
$\bm{E}^{s^{\mu}}$ is the (virtual) driving force
for the spin current, i.e.,

$\langle\bm{j}^{c}\rangle_{J_{K}=0}
= \frac{e^{2}n_{e}\tau}{m^{*}}\bm{E}$,
$\langle\bm{j}^{s^{\mu}}\rangle_{J_{K}=0} 
= \frac{e_{s}^{2}n_{e}\tau}{m^{*}}\bm{E}^{s^{\mu}}$,
where $n_{e} = k_{F}^{3}/(3\pi)^{2}$
and $k_F$ is the Fermi momentum.

Under the above assumption
the equations of motion of discrete spins are obtained
by using the relation 
$\bm{S}(\bm{r})
\leftrightarrow
\sum_{n}\delta(\bm{r}-\bm{r}_{n})\bm{S}_{\bm{r}_{n}}$.
\begin{eqnarray}
\frac{\partial\tilde{\bm{S}}_{i}}{\partial \tilde{t}}
&\cong& 
\gamma^2
F^{(1)}_{0}(0)\tilde{\bm{S}}_{i}
\times\frac{\partial\tilde{\bm{S}}_{i}}{\partial \tilde{t}}
\nonumber\\
&&
+\sum_{j\neq i}
\Biggl[
\left[F^{(0)}_{0}(\bm{r}_{ij})+F^{(0)}_{\mathrm{cc}}(\bm{r}_{ij})\right]
\tilde{\bm{S}}_{i}\times\tilde{\bm{S}}_{j}
\nonumber\\
&&\quad
-\tilde{\bm{S}}_{i}\times\left[\bm{F}^{(0)}_{\mathrm{sc}}(\bm{r}_{ij})\times\tilde{\bm{S}}_{j}\right]
\nonumber\\
&&\quad
+\gamma^2
\left[F^{(1)}_{0}(\bm{r}_{ij})+F^{(1)}_{\mathrm{cc}}(\bm{r}_{ij})\right]
\tilde{\bm{S}}_{i}\times\frac{\partial\tilde{\bm{S}}_{j}}{\partial \tilde{t}}
\nonumber\\
&&\quad
-\gamma^2
\tilde{\bm{S}}_{i}\times
\left[\bm{F}^{(1)}_{\mathrm{sc}}(\bm{r}_{ij})\times
\frac{\partial\tilde{\bm{S}}_{j}}{\partial \tilde{t}}\right]
\Biggr],
\label{eq:EOM}
\end{eqnarray}
where $\bm{r}_{ij}=\bm{r}_{i}-\bm{r}_{j}$
and  $\tilde{\bm{S}}_{i}=\bm{S}_{\bm{r}_{i}}/S$.
The time scale is normalized as 
$\tilde{t} = \gamma^{2}\epsilon_{F}t$ where
$\gamma = \sqrt{S}v_{\mathrm{uc}}n_{e}J_{K}/(2\epsilon_{F})$.
In the above equation of motion, we have 
retained the time derivatives up to $n\le 1$.
The detailed forms of dimensionless functions
$F^{(n)}_{0,\mathrm{cc}}(\bm{r})$ and $\bm{F}^{(n)}_{\mathrm{sc}}(\bm{r})$
for the $n$-th time-derivatives
are given elsewhere \cite{OnodaLong}. 
$F^{(0)}_{0}(\bm{r})$ is 
the Ruderman-Kittel-Kasuya-Yosida (RKKY) range function,
$F^{(1)}_{0}(\bm{r})
= -\frac{9\pi}{4}\cdot
\frac{1-\cos(2k_{F}r)}{(2k_{F}r)^2}$,
$F^{(n)}_{\mathrm{cc}}(\bm{r})\propto \bm{r}\cdot\langle\bm{j}^{c}\rangle_{J_{K}=0}$
and $F^{(n),\mu}_{\mathrm{sc}}(\bm{r})\propto 
\bm{r}\cdot\langle\bm{j}^{s^{\mu}}\rangle_{J_{K}=0}$.
It is noted that at least two spins are needed in order to feel
the flow of the charge/spin currents.
This results is roughly consistent with 
the result obtained by Barnes for the dynamical RKKY problem
\cite{Barnes}.
It can be shown that 
the above equations of motion is valid
in the present situation $\gamma^{2} \ll 1$
and the magnetic field is not too strong \cite{OnodaLong}.

$F^{(0)}_{0}(\bm{r})$-term is the intrinsic torque
between the localized spins due to the RKKY interaction.
$F^{(1)}_{0}(0)$-term represents the Gibert damping
due to the Fermi bath of the conduction electrons,
and $F^{(1)}_{0}(\bm{r})$-term ($r\neq0$) 
works as a non-local damping/puming \cite{Barnes,RikitakeImamura}.
$F^{(0)}_{\mathrm{cc}}(\bm{r})$-/$\bm{F}^{(0)}_{\mathrm{sc}}(\bm{r})$-term
is the intrinsic torque due to the charge/spin currents.
Depending on the separation of spins
and the flow of charge/spin currents,
$F^{(0)}_{\mathrm{cc}}(\bm{r})$-term renormalizes the original toruqe term
but $\bm{F}^{(0)}_{\mathrm{sc}}(\bm{r})$-term
introduces a torque term with a different symmetry,
because the spin current is not a vector but a tensor.
As we shall see later, this term introduces the spin anisotropy
in the direction of the spin component of the spin current.
$F^{(1)}_{\mathrm{cc}}(\bm{r})$-/$\bm{F}^{(1)}_{\mathrm{sc}}(\bm{r})$-term
is the correction to the non-local damping/pumping term
due to the charge/spin currents.
$\bm{F}^{(1)}_{\mathrm{sc}}(\bm{r})$-term has a different symmetry
from that of the original term and leads to
a spin-anisotropic damping/pumping.
\\
\noindent
{\it Dynamics of two-spin system}--
Now we consider the dynamics of the two-spin system in detail.
For the numerical study, we normalize the charge/spin currents as
\begin{eqnarray}
\bm{J}^{n} 
&=& \frac{\lambda_{F}^{2}}{2e\epsilon_{F}}
\langle\bm{j}^{c}\rangle_{J_{K}=0},
\quad
\bm{J}^{s^{\mu}} 
= \frac{\lambda_{F}^{2}}{2e_{s}\epsilon_{F}}
\langle\bm{j}^{s^{\mu}}\rangle_{J_{K}=0},
\end{eqnarray}
where $\lambda_{F} = 2\pi/k_{F}$,
and, take the forms of them as
$\bm{J}^{n} = J^{n} \bm{j}_{0}$ and
$\bm{J}^{s^{\mu}} = J^{s}n^{\mu}_{s}\bm{j}_{0}$,
where $\bm{j}_{0}$ and $\bm{n}_{s}$ are the unit vectors 
representing the directions of the charge/spin-current flow
and the spin-current polarization, respectively.
The unit of charge current, 
$2e\epsilon_{F}/\lambda_{F}^{2}$,
is estimated as $\sim 10^{5}$~Acm$^{-2}$ for 
a semiconductor with $n_{e}\sim 10^{18}$~cm$^{-3}$ 
and $\epsilon_{F} \sim 1$~meV,
and as $\sim 10^{10}$~Acm$^{-2}$ for 
a metal with $n_{e}\sim 10^{21}$~cm$^{-3}$ 
and $\epsilon_{F} \sim 1$~eV.
The unit of time, $1/(\gamma^2\epsilon_F)$,
is estimated as $10\times$(1 meV/$\epsilon_{F}$)~ps
for $\gamma =0.05$.
We put the two spins along $\bm{j}_{0}$
which is set to be the $x$-direction.
$\bm{n}_{s}$ is set to be the $z$-direction.

\begin{figure}[hbt]
\includegraphics[scale=0.3]{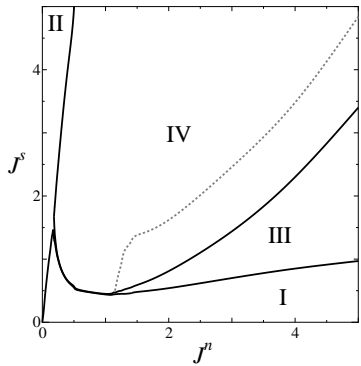}
\caption{
Phase diagram of the dynamics of the two-spin system
in the plane of charge current $J^n$ and spin current $J^s$. 
} 
\label{fig:phase}
\end{figure}
\begin{figure}[hbt]
\includegraphics[scale=0.3]{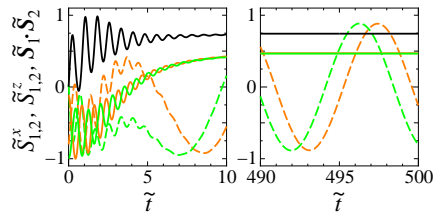}
\caption{
Example of the motion in the region II.
$J^{n}=0.1$ and $J^{s} = 2.0$. For the meaning of each curves, see the text.
} 
\label{fig:motion-II}
\end{figure}
\begin{figure}[hbt]
\includegraphics[scale=0.3]{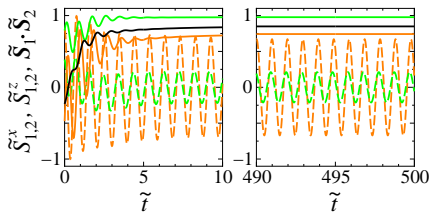}
\caption{
Example of the motion in the region III.
$J^{n}=2.0$ and $J^{s} = 0.6$.
} 
\label{fig:motion-III}
\end{figure}
\begin{figure}[hbt]
\includegraphics[scale=0.3]{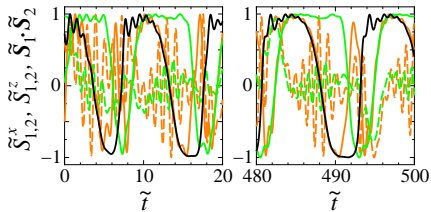}
\caption{
Example of the motion in the region IV.
$J^{n}=1.0$ and $J^{s} = 1.0$.
} 
\label{fig:motion-IV}
\end{figure}

Fig.~\ref{fig:phase} show the `` phase diagram'' for 
the dynamics of the two-spin system
with $k_{F}r = 1.0$ and $\gamma^{2}=0.05$.
Figs.~\ref{fig:motion-II}-\ref{fig:motion-IV}
show examples of motions in the regions II-IV respectively.
The black line represent the inner product of the two spins
$\tilde{\bm{S}}_{1}\cdot\tilde{\bm{S}}_{2}$.
The solid lines of green and orange
are the $z$-components of upstream and downstream spins respectively,
i.e., $\tilde{S}^{z}_{1}$ and $\tilde{S}^{z}_{2}$.
The dashed lines of green and orange
are the $x$-components of upstream and downstream spins respectively,
i.e., $\tilde{S}^{x}_{1}$ and $\tilde{S}^{x}_{2}$.

In region I, the two spins reach the final static collinear configuration
with the easy-axis anisotropy in the direction of $\bm{n}_{s}$.
In the special case of $J^{s}=0$, i.e., the pure charge current,
the static collinear configuration is realized with no spin anisotropy.
This region contains the cases with
any amount of the weakly spin-polarized charge current and
a small amount of the perfelctly spin-polarized charge current.
In region II and III, the spins show the periodic motions
in which the angle between the two spins
and the $z$-component of each spin takes a constant value.
The weakly-charged spin current belongs to region II,
which is our main concern.
An example of this case is given in Fig.~\ref{fig:motion-II}, 
where the lines for $S^{z}_{1}$ (green solid) and $S^{z}_{2}$ (orange solid) 
are almost overlapped.
When $|J^{n}|$ is sufficiently smaller than $|J^{s}|$
the angle $\theta_{12}$ between the two spins
is a increasing (decreasing) function of $|J^{s}|$ ($|J^{n}|$).
The $z$-componet of a spin
is a increasing (decreasing) function of $|J^{n}|$ ($|J^{s}|$).
The frequency of motion increases with both of $|J^{n}|$ and $|J^{s}|$.
In the case with $J^{n}=0$, i.e., the pure spin current,
the static coplanar configuration 
with the hard-axis parallel (or the easy-plane perpendicular)
to $\bm{n}_{s}$ is realized
and the angle between the two spins, $\theta_{12}$,
is determined by the magnitude of $J^{s}$ as
$|\tan\theta_{12}|=|\bm{F}^{(0)}_{\mathrm{sc}}(\bm{r}_{12})/F^{(0)}_{0}(\bm{r}_{12})|$.
Fig.~\ref{fig:motion-III} shows another type of a simple motion which occurs 
in region III.
In region IV observed the chaotic or complicated motion 
in which both the angle between the two spins
and the $z$-component of each spin are changing in motion and do not reach a 
steady state. This region contains the cases with
a large amount of the strongly spin-polarized charge current
and a large amount of the strongly charged spin current.
Although we observe various types of motions in this region,
any sharp boundaries between them are not identified
at least in the present simulation.
We shall not undertake their precise classification.
However, a relatively clear change can be observed
around the dotted line in Fig.~\ref{fig:phase}.
Under the dotted line in this region,
the upstream spin slowly precesses around $\bm{n}_{s}$ with 
a cycloidal trajectory and the downstream spin quickly precesses around 
the upstream spin.
Deep in the region IV above the dotted line,
the motion becomes chaotic as shown in 
Fig.~\ref{fig:motion-IV}.
Note that a large amount of 
strongly spin-polarized charge current
leads to this class of motion, where the 
strong coupling theories ( for $\gamma^{2}\gg1$ )
\cite{Bazaliy,Fernandez,Shibata}
predict the instability of the ferromagnetic order.
Some remarks are in order here.
First the dynamics highly depends on the initial condition
in the vicinity of the boundary between the regions II and IV.
Secondly there is a very narrow region between the regions I and IV,
while the regions I and IV appear to touch each other.
In this narrow region,
we observe the motions similar to those in the regions II and III.
Thus, the regions II and III may be 
connected through this region.\\

\noindent
{\it Discussion}--
The above consideration on the case of pure spin current
can be easily generalized to the one-dimensional 
periodic array of spins along the direction of $\bm{j}_{0}$.
The equilibrium interaction $F^{(0)}_0(\bm{a}_{0})$
between the nearest neighbor spins with the separation
$\bm{a}_{0}$ is assumed to be ferromagnetic. 
Even in the case with $J^{s}\neq 0$ but $J^{n}=0$, 
we can construct an energy functional $E_{\mathrm{eff}}$
which should be minimized for the spin configuration of
final steady state.
\begin{eqnarray}
E_{\mathrm{eff}}
&=& -\sum_{i<j}
\left[F^{(0)}_{0}(\bm{r}_{ij})\bm{S}_{i}\cdot\bm{S}_{j}
+\bm{F}^{(0)}_{\mathrm{sc}}(\bm{r}_{ij})\cdot[\bm{S}_{i}\times\bm{S}_{j}]
\right].
\nonumber\\
\end{eqnarray}
This suggests that even the dissipative 
transport spin current works on the localized spins
in the same way as the equilibrium spin current does,
while it is not the case for the charge current.
In equilibrium, i.e., $J^{n} = J^{s} = 0$, 
the spins align ferromagnetically, and the direction of alignment
is arbitrary because there is no spin anisotropy in our model.
When $J^{s}$ is introduced, the spins feel 
Dzyaloshinsky-Moriya-like interaction in addition
to the ferromagnetic one.
For simplicity, we shall take into account 
only the nearest neighbor interaction
and set $\bm{n}_{s}$ to be in the positive $z$-direction.
Then, the variational energy including the lowest energy state is given by 
$E_{q} = -F^{(0)}_{0}(\bm{a}_{0})\cos(qa_{0})
+F^{(0),z}_{\mathrm{sc}}(\bm{a}_{0})\sin(qa_{0})$.
The lowest energy state is realized 
by the real-space configuration 
$\bm{S}_{i}/S = [\cos(q_{0}r_{i}+\phi_{0}), \sin(q_{0}r_{i}+\phi_{0}), 0]$
where 
$\tan(q_{0}a_{0}) = -F^{(0),z}_{\mathrm{sc}}(\bm{a}_{0})/F^{(0)}_{0}(\bm{a}_{0})$
and $\phi_{0}$ is an arbitrary constant.
The spins is expected to show the helical order
and the hard-axis anisotropy parallel
(or the easy-plane anisotropy perpendicular) to $\bm{n}_{s}$
is introduced by the pure spin current.
The pitch angle, $|q_{0}a_{0}|$, of the helix increases with
the magnitude of $|J^{s}|$,
because $|\bm{F}^{(0)}_{\mathrm{sc}}(\bm{a}_{0})|\propto |J^{s}|$.
Hence this helical order
is realized in the steady state with the pure spin current,
this might be utilized as the spin current detector.
In other cases than discussed above,
we must consider the dynamical equations of motion directly.
However, the previous study of the two-spin system
gives us the following insights.
When the small amount of $J^{n}$ is introduced to the pure spin current,
the spins arise from the coplanar helical order in the easy plane
and show the periodic motion around $\bm{n}_{s}$
as in the motion of the two spins in the region II.
In the case with a small amount of the weakly spin-polarized current 
which belongs to the region I,
the spins align ferromagnetically as in equilibrium,
but in this case the easy-axis anisotropy parallel to  $\bm{n}_{s}$
is introduced by the spin-polarized current.
However, for a large amount of the strongly spin-polarized charge current
belonging to the region IV,
the instability of ferromagnetic order would be observed.
The details of the dynamical motion of multi-spin system
will be reported elsewhere.

In conclusion, we have derived the equations of motion
of localized spins weakly coupled to the conduction electrons
with charge/spin currents.
The dynamics of localized spins are affected
significantly by these charge and spin currents
of conduction electrons, and the dynamics of the 
two-spin system has been studied in depth by numerical 
calculation. The phase diagram for the dynamics in the plan 
of charge and spin currents are divided into 4 regions,
where the steady state, periodic motion, and even chaotic
motion are observed. 

The authors thank S.~E.~Barnes, M.~Maekawa,
J.~Ieda, H.~Imamura, S.~Takahashi, G.~E.~W.~Bauer, 
A.~V.~Balatsky, and G.~Tatara for fruitful discussions.
This work is financially supported by NAREGI Grant,
Grant-in-Aids from the Ministry of Education,
Culture, Sports, Science and Technology of Japan.

\end{document}